\documentclass[fleqn,12pt,twoside]{article}
\usepackage{espcrc1}
\usepackage{epsfig}

\usepackage{graphicx}
\usepackage[figuresright]{rotating}

\newcommand{\AmS}{{\protect\the\textfont2
  A\kern-.1667em\lower.5ex\hbox{M}\kern-.125emS}}

\title{Experimental review of unpolarised nucleon structure functions}

\author{Vladimir Chekelian (Shekelyan)\address[MCSD]{Max-Planck-Institut f\"ur Physik (Werner-Heisenberg-Institut) \\
F\"ohringer Ring 6, 80805 M\"unchen, Germany \\
E-mail: shekeln@mppmu.mpg.de}}
       
\begin{document}

% typeset front matter
\maketitle

\begin{abstract}
Recent results are reviewed on unpolarised structure functions
from fixed target experiments at JLAB, NuTeV and 
from the HERA $ep$ collider experiments H1 and ZEUS.

\end{abstract}

\section{Introduction}

The study of the structure of hadrons has been a powerful means
for establishing and testing of the theory of strong interactions
and for the determination of the partonic content of the nucleon.
In 1969 the SLAC-MIT collaboration~\cite{slac} 
observed the scaling behaviour~\cite{bjorken} of the proton structure function 
in deep inelastic electron-proton scattering (DIS).
This observation established the quark-parton model (QPM)
as a valid framework for the interpretation of DIS data and that 
different partonic processes can be expressed 
in terms of universal parton densities.
The smallness of $R=\sigma_L/\sigma_T$, the ratio of 
the cross sections from longitudinally and transversally 
polarised virtual photon scattering measured in DIS~\cite{spin},
provided the first evidence of the spin-1/2 nature of the partons.
In 1974 in $\mu N$ interactions~\cite{violations} and in subsequent neutrino
and muon-nucleon scattering experiments, the observation of
scaling violation and the identification of partons as quarks and gluons has
confirmed the field theory of quarks and gluons and their strong interactions, 
Quantum Chromodynamics (QCD).  

The structure functions in QCD
are defined as a convolution of the universal
parton momentum distributions inside the proton
and coefficient functions, which contain information
about the exchanged boson-parton interaction.
At sufficiently large four-momentum transfer squared, $Q^2$,
when the strong coupling $\alpha_s$ is small,
a perturbative technique
is applicable for QCD calculations of the coefficient
and splitting functions. The latter
represent the probability
of a parton to emit another parton.
Expressions for structure functions are determined by
convolution integrals of appropriate sums over
the densities of quarks of different flavours and gluons, which predict
a logarithmic $Q^2$ dependence (evolution) of the structure functions.
The perturbative QCD (pQCD)  calculations are well established
within the DGLAP~\cite{dglap} formalism
to next-to-leading (NLO) order in the strong coupling
and have recently been extended to next-to-next-to-leading
(NNLO) order~\cite{sm}.

\begin{figure}[htb]
%\vspace*{3.6cm}
 \begin{picture}(170,200)(0.,0.)
 \put(0.,-10.){\epsfig{file=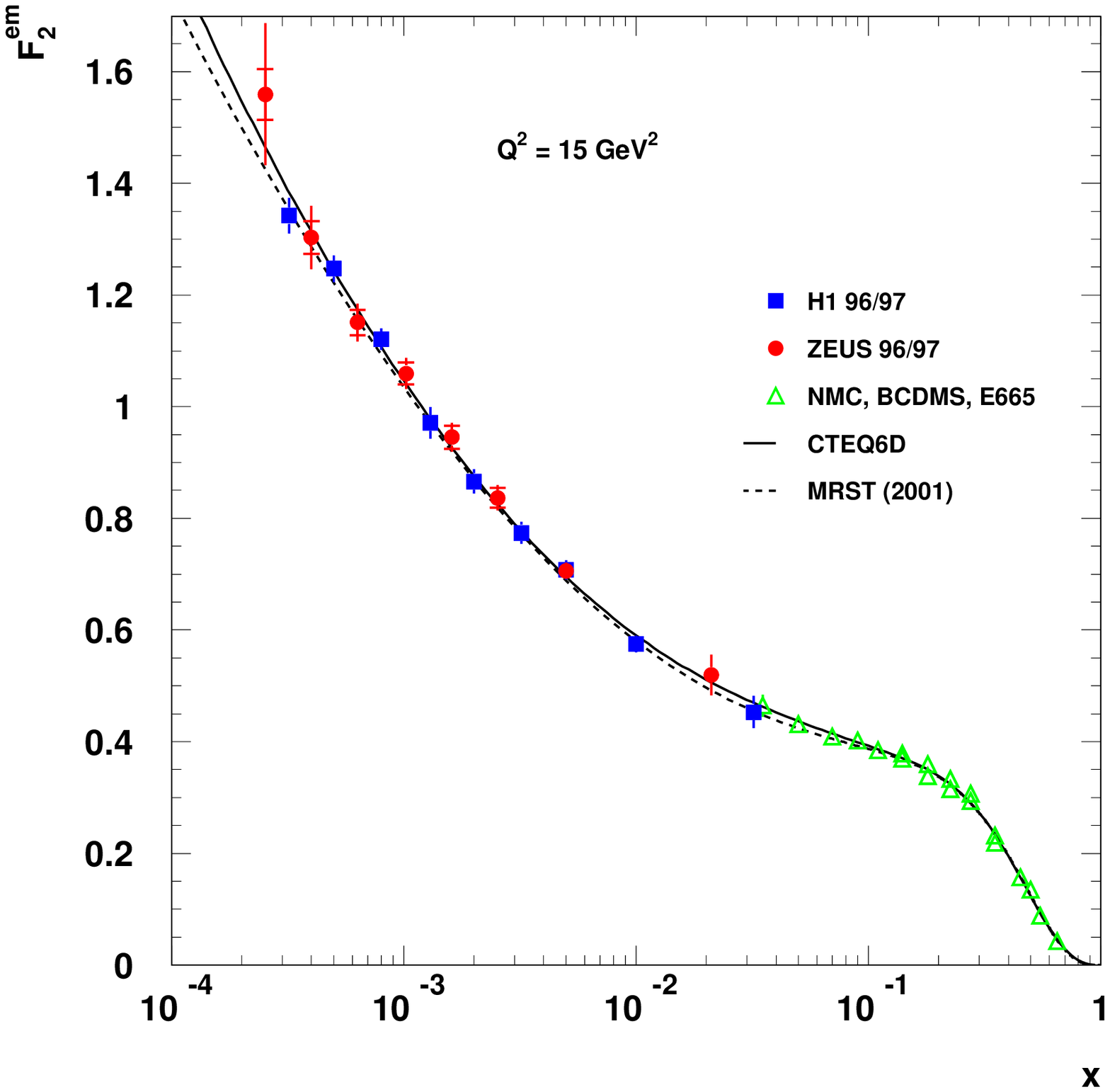,height=8cm}}
%            bbllx=0.,bblly=0.,bburx=540.,bbury=760.,height=9cm}}
 \end{picture}
 \begin{picture}(170,200)(0.,0.)
 \put(50.,-30.){\epsfig{file=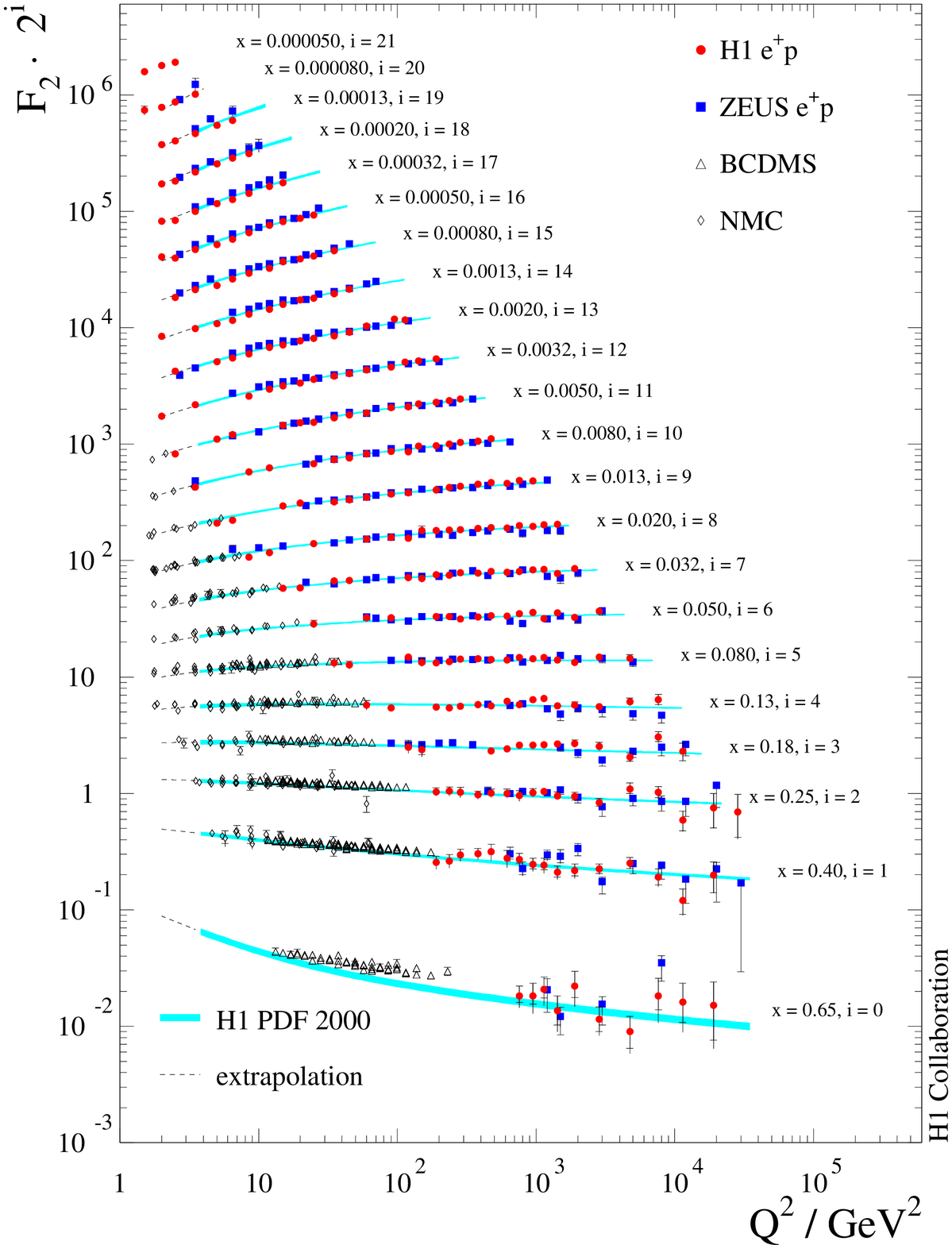,height=9cm}}
 \end{picture}
\caption{
Proton structure function $F_2(x,Q^2)$ measured at HERA and 
in fixed target experiments as a function of $x$ 
at $Q^2=15~$GeV${^2}$ (left) and as a function of $Q^2$  
for different $x$ values (right). Curves represent different NLO QCD fits.}
\label{f2}
\end{figure}

In 2004 D. Gross, D. Politzer and F. Wilczek 
have been awarded the Nobel Prize for the discovery of asymptotic freedom.
F. Wilczek in his brief commentary~\cite{wilczek}
on the QCD foundational papers wrote
that the ``most dramatic'' experimental consequence  
``regarding the pointwise evolution of structure functions'', namely
``that the proton viewed at ever higher resolution would appear more and more
as field energy (soft glue), was only clearly verified
at HERA twenty years later''. 
Indeed, the first measurements by the HERA electron proton collider
experiments H1 and ZEUS in 1992~\cite{h1f2-92,zeusf2-92} 
revealed a steep rise of the proton structure
function towards small $x$, the fraction of proton momentum 
carried by the parton.
The present state of the art of 
these measurements~\cite{klein,h1lowq2,zeuslowq2,h1highq2,zeushighq2}
is shown in Figure~\ref{f2} as a function of $x$ 
at $Q^2=15~$GeV${^2}$ (left figure) and as a function of $Q^2$ at 
different $x$ values (right figure).
 
Partons, i.e. quarks and gluons, enter differently
into different structure functions.
In deep inelastic scattering 
the cross section of the neutral or charged current (NC or CC) process
can be expressed in terms of three structure functions.
For example for neutral current process
mediated by $\gamma$ or $Z^{\rm o}$-boson exchange,
the expression is
\begin{equation}
  \frac{d^2\sigma^{e^{\pm}p \rightarrow e^{\pm}X}}{dx dQ^2}
  =\frac{2\pi\alpha^2}{Q^4x}
  \left[Y_+F_2(x,Q^2)-y^2 F_L(x,Q^2)\mp Y_-xF_3(x,Q^2)\right],
\label{dsignc}
\end{equation}
where $y= Q^2/xs$ is the inelasticity,
$s$ is the center of mass energy squared,
$\alpha$ is the fine structure constant
and $Y_{\pm}=1{\pm}(1-y)^2$.
The dominant contribution to the cross section is due to
the proton structure function $F_2$, which,
in the framework of QPM, is related to the sum of the
proton momentum fractions carried by the quarks and antiquarks
in the proton weighted by the quark charges squared.
The longitudinal structure function $ F_{L}$,
vanishing in QPM, is
directly sensitive to the gluon momentum distribution in the proton.
The structure function $ xF_{3}$ depends on the valence quarks
and is sizable only at large $Q^2$,
when $Q^2$ is comparable with the $Z^{\rm o}$-boson
mass squared.

In this paper the recent JLab results on the
transverse and longitudinal structure functions 
in the nucleon resonance region,
the NuTeV measurements of the isoscalar structure
functions $F_2(x,Q^2)$ and $xF_3(x,Q^2)$ in $\nu Fe$ interactions
and the HERA results on structure functions are reviewed 
as presented at the BARYONS 2004 conference.
The HERA results include analyses of the inclusive NC and CC $e^{\pm}p$
cross section measurements in the framework of perturbative NLO QCD,
determination of the quark and gluon distributions
inside the proton, results on the strong coupling 
obtained from inclusive and jet data
and also measurements of the charm and beauty contributions to $F_2$.

%%%%%%%%%%%%%%%%%%%%%%%%%%%%%%%%%%%%%%%%%%%%%%%%%%%%%%%%%%%%%%%%%%%%
%%%%%%%%%%%%%%%%%%%%%%%%%%%%%%%%%%%%%%%%%%%%%%%%%%%%%%%%%%%%%%%%%%%%
\section{Recent results from fixed target experiments}

\begin{figure}[htb]
 \begin{picture}(170,160)(0.,0.)
 \put(5.,-20.){\epsfig{file=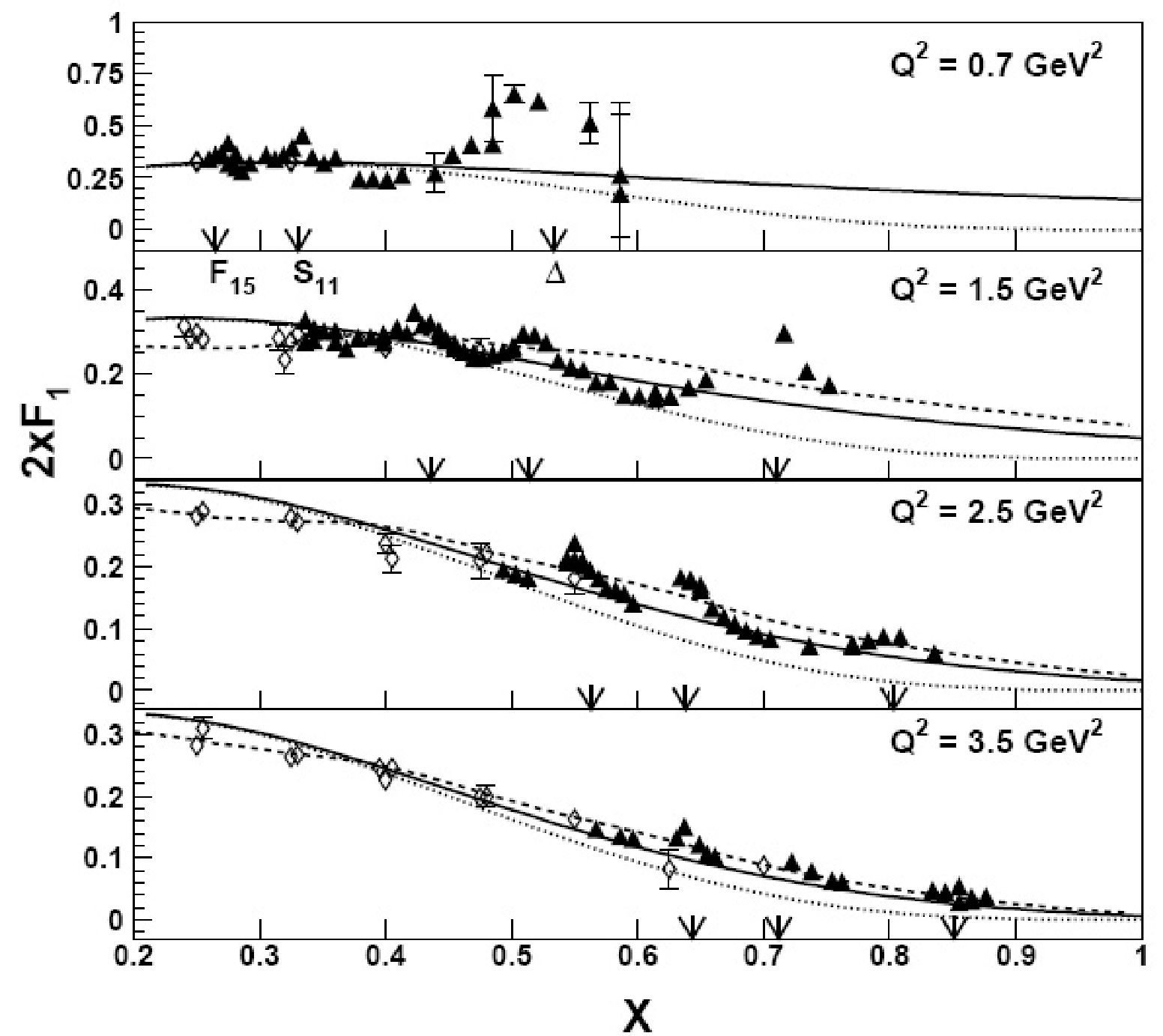,height=6.8cm}}
 \end{picture}
 \begin{picture}(170,160)(0.,0.)
 \put(50.,-20.){\epsfig{file=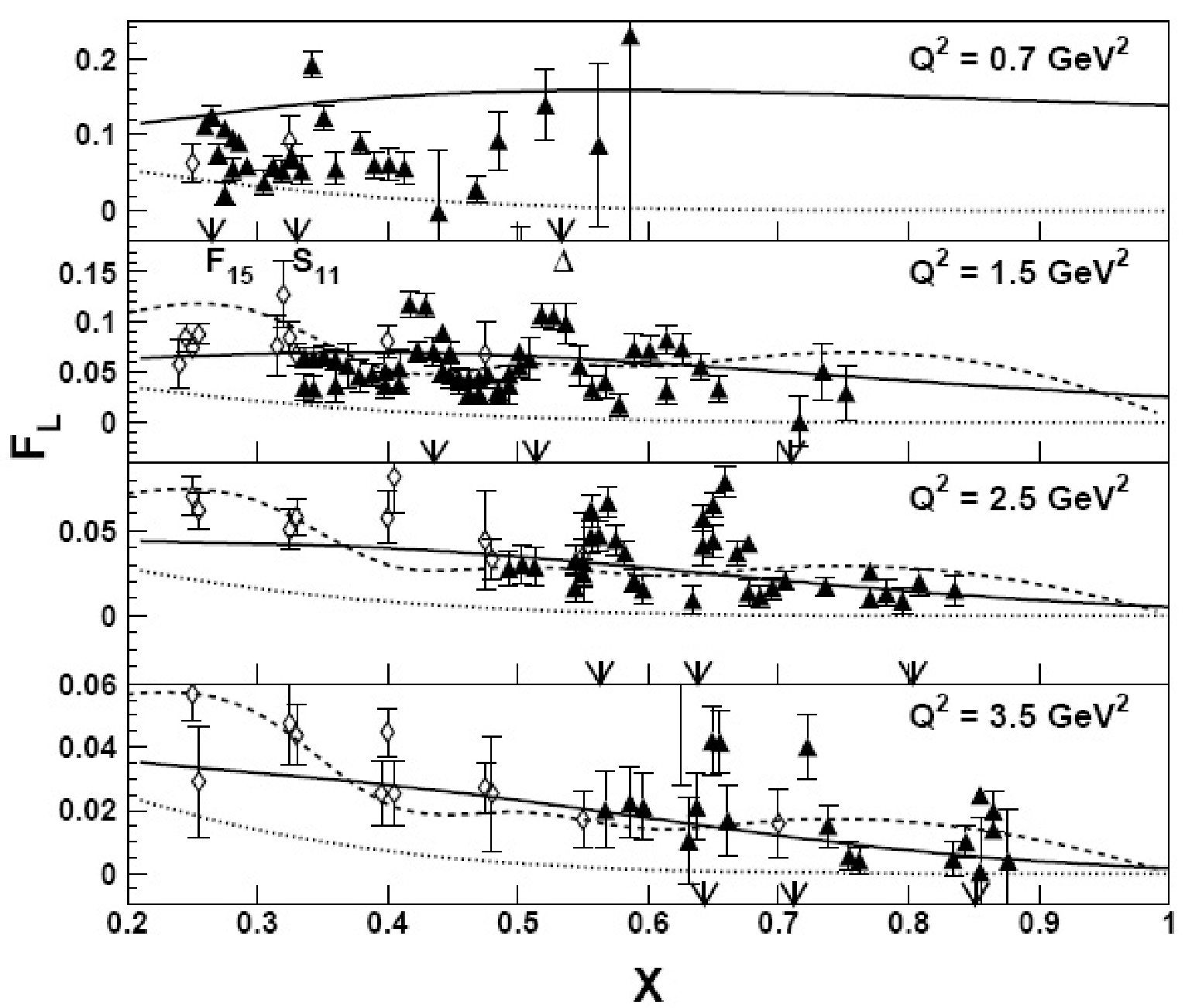,height=6.8cm}}
 \end{picture}
\caption{The transverse ($2xF_1$) and longitudinal ($F_L$) 
structure functions from the E94-110 experiment (triangles)
compared with the SLAC measurements (diamonds) and
the NNLO QCD fits from Alekhin~\cite{alekhin-jlab} (dashed line) 
and MRST~\cite{mrst-jlab} (solid line)
with, and MRST (dotted line) without target mass effects included.
The prominent resonance mass regions are indicated by arrows. }    
\label{jlab}
\end{figure}

The Jefferson Lab experiment E94-110
measured inclusive scattering of 
unpolarised electrons off a hydrogen target 
in the nucleon resonance region,
in the range $0.2<Q^2<5.5$~GeV$^2$~\cite{jlab}.
The longitudinal-transverse separation
allowed the proton structure functions 
$F_2=(2xF_1+F_L)/(1+4M^2x^2/Q^2)$, 
$F_1$ (purely transverse) and $F_L$ (purely longitudinal)
to be extracted independently. Here, $M$ is the proton mass.
The measurements of $2xF_1$ and $F_L$ are shown in Figure~\ref{jlab}
as a function of $x$ for different $Q^2$ values.
The longitudinal component is found to be significant,
both structure functions show resonant mass structures
which oscillate around the lines corresponding to QCD fits. 
At $Q^2\ge 1.5$~GeV$^2$ the average 
$x$ dependence  of the resonance region is well described.
The observed scaling relationship between resonance electroproduction
and deep inelastic scattering,
termed quark-hadron (Bloom-Gilman) duality~\cite{duality},
suggests a common origin for both kinematic regimes and brings additional 
information on the transition from soft to hard QCD.
These data represent the first observation of duality
in the separated transverse and longitudinal structure functions.

\begin{figure}[htb]
 \begin{picture}(170,200)(0.,0.)
 \put(60.,-25.){\epsfig{file=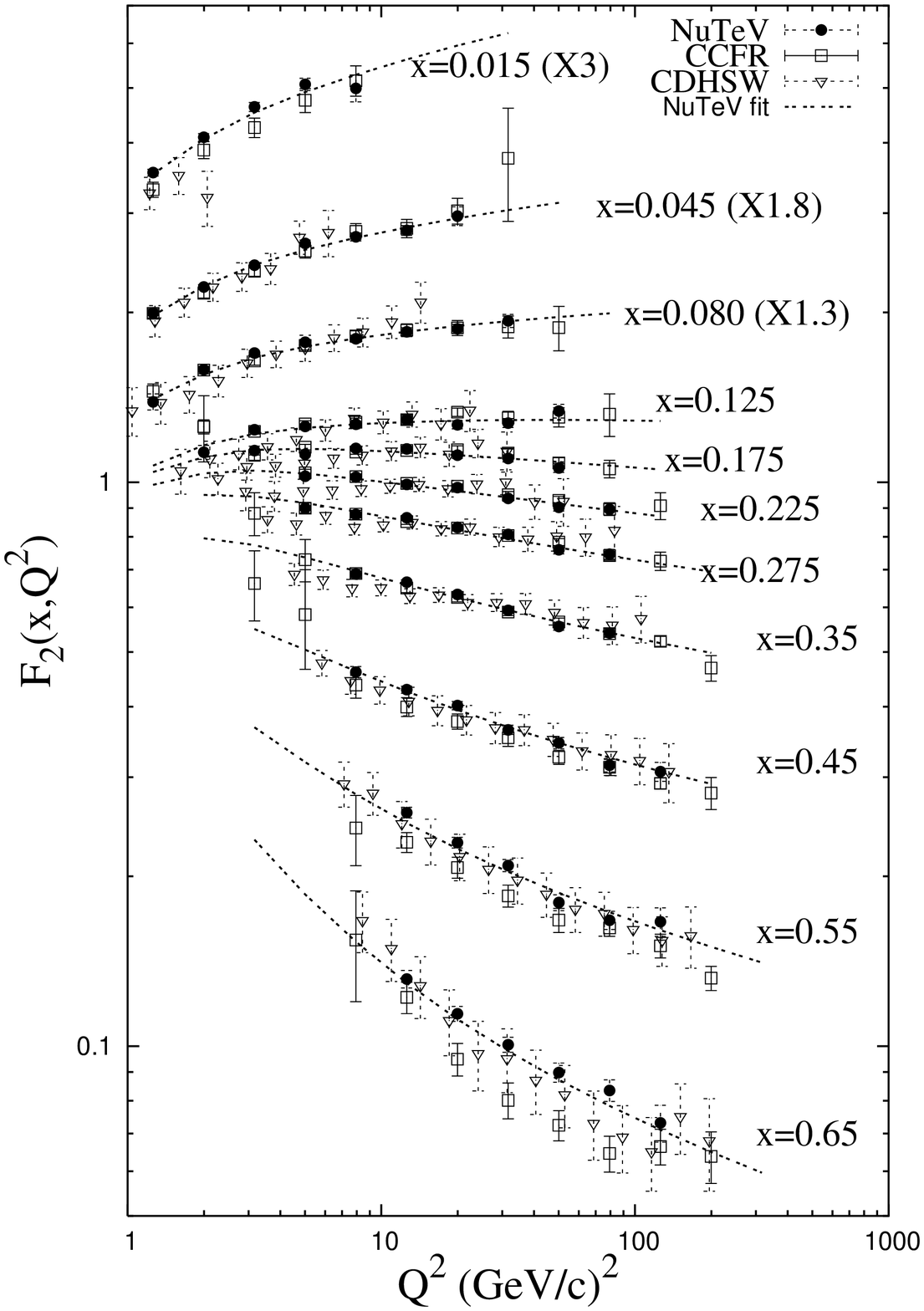,height=8cm}}
 \put(220.,-25.){\epsfig{file=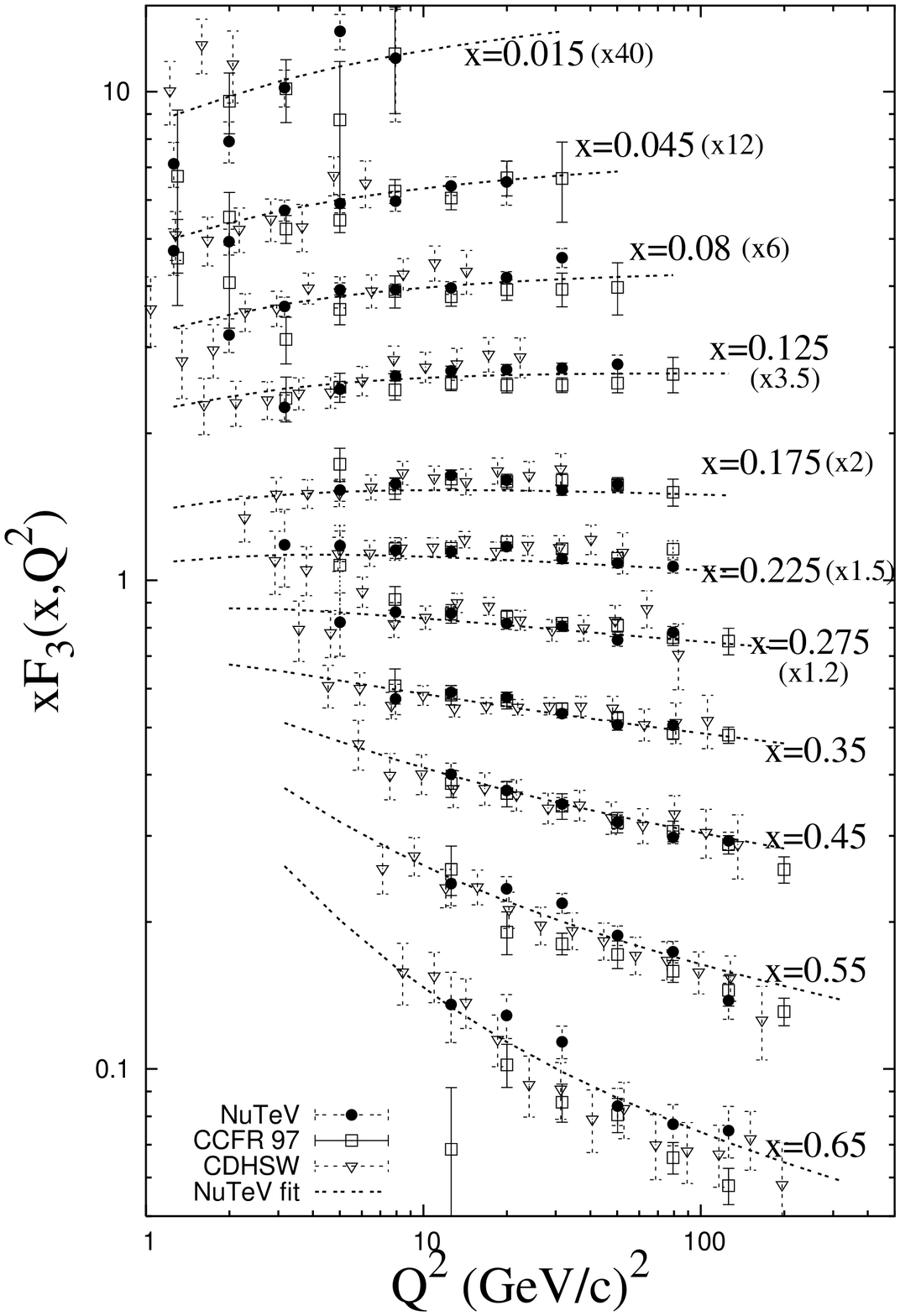,height=8cm}}
 \end{picture}
\caption{
The NuTeV measurements of $F_2(x,Q^2)$ and
$xF_3(x,Q^2)$ compared with the data from CCFR and CDHSW. 
The curve represents the fit to the NuTeV data. }
\label{nutev}
\end{figure}

The NuTeV experiment has obtained 
high statistics samples of neutrino and antineutrino 
CC interactions
using the high energy sign-selected neutrino beam at Fermilab. 
Isoscalar neutrino-iron structure
functions $F_2(x,Q^2)$ and $xF_3(x,Q^2)$~\cite{nutev}
determined by fitting the $y$ dependence of the sum and the difference of
$\nu$ and $\bar{\nu}$ cross sections are shown in Figure~\ref{nutev}.
The measurements benefit from
the precise determination of muon and hadron energy
scales (0.7\% and 0.43\%, respectively)
obtained using continuous calibration beam running concurrently 
with the data taking. 
At intermediate $x$ ($0.015 < x < 0.5$) both $F_2(x,Q^2)$ and 
$xF_3(x,Q^2)$ are in 
good agreement with the CCFR results which are also
shown in Figure~\ref{nutev}.
At lowest $x$ ($x=0.015$) and in the high $x$ region ($x > 0.5$)
the NuTeV data are
systematically above the CCFR results by $\approx 3$~\% and
by up to $\approx 20$~\%, respectively. 
The new measurements have improved systematic precision,
and they have expanded into the previously unaccessible kinematic range 
of high inelasticity y.

%%%%%%%%%%%%%%%%%%%%%%%%%%%%%%%%%%%%%%%%%%%%%%%%%%%%%%%%%%%%%%%%%%%%
%%%%%%%%%%%%%%%%%%%%%%%%%%%%%%%%%%%%%%%%%%%%%%%%%%%%%%%%%%%%%%%%%%%%
\section{HERA results}

\begin{figure}[htb]
 \begin{picture}(170,190)(0.,0.)
 \put(10.,-10.){\epsfig{file=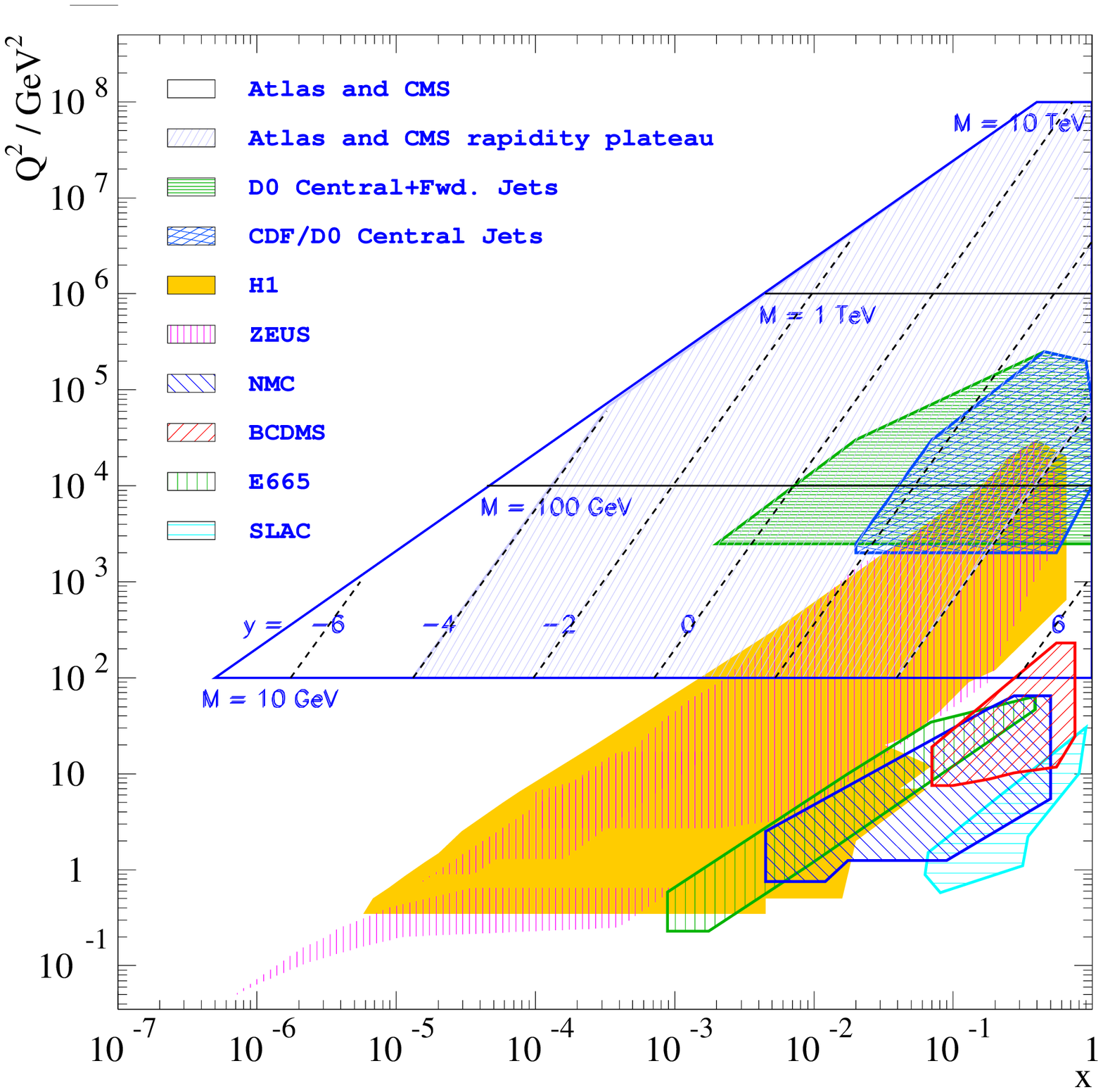,height=7.2cm}}
 \end{picture}
 \begin{picture}(170,190)(0.,0.)
 \put(40.,-40.){\epsfig{file=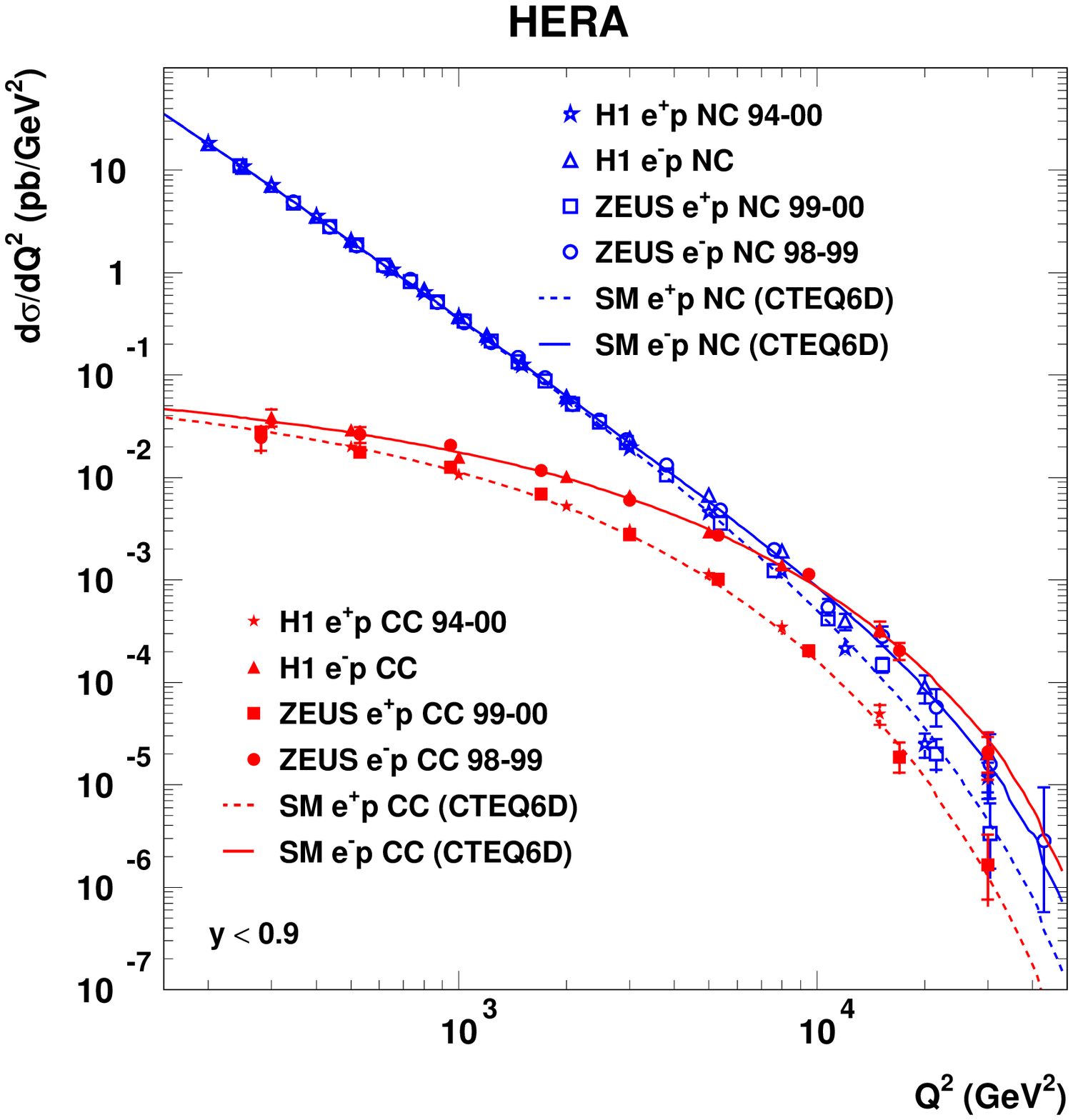,height=9cm}}
 \end{picture}
\caption{
Kinematic reach at the HERA collider and in the fixed target experiments 
together with the regions of jet production in $pp$ scattering (left).
The H1 and ZEUS measurements of the NC and CC $e^{\pm}p$ cross sections
as a function of $Q^2$ (right).}
\label{kinplane}
\end{figure}

The kinematic coverage of the HERA structure
function measurements in the $x$-$Q^2$ plane
is shown in Figure~\ref{kinplane} (left). 
HERA extends the phase space of the previous
fixed target measurements (also shown in the figure) 
by 2 orders of magnitude both in $x$ and $Q^2$. 
The full HERA range in $x$ is essential for predictions 
for the LHC collider.

At the HERA collider, the incident electron (or positron) energy is 
27.6~GeV and the proton energy is 920 GeV (820 GeV till 1997),
which correspond to a center of mass energy of $319~(301)$~GeV.  
In the first phase of data taking, HERA I,
the  H1 and ZEUS experiments each collected about
100 pb{$^{-1}$} of integrated luminosity in the $e^+p$ mode and
about 15 pb{$^{-1}$} in the $e^-p$ mode.
The H1 and ZEUS measurements 
of the single differential NC and CC $e^{\pm}p$ cross sections 
$d\sigma/dQ^2$~\cite{h1highq2,zeushighq2,zeuscc,h1elec,zeuselec}
are summarised in Figure~\ref{kinplane} (right).
At low $Q^2\approx~100$~GeV${^2}$ the cross section of
the CC process mediated by the $W$ boson,
is smaller by 3 orders of magnitude than that of the NC process,
due to the different propagator terms.
At high $Q^2\approx M_Z^2, M_W^2$
the cross section measurements are approaching each other
demonstrating the unification of the weak and electromagnetic forces.
From a comparison of the NC measurements at highest $Q^2$ with
the Standard Model expectation, 
a limit on the quark radius of $\approx10^{-18}$~m is obtained,
proving a pointlike behaviour of quarks
down to about 1/1000 of the proton radius. 

\subsection{Partonic structure of the proton}

The measurements of the full set of NC and CC double differential
$e^{\pm}p$ cross sections at HERA
allow comprehensive QCD analyses
to determine the quark and gluon distributions
inside the proton and the strong coupling constant $\alpha_s(M_Z^2)$.
%and to explore the limits of pQCD application, for example at small $x$.

\begin{figure}[htb]
 \begin{picture}(170,190)(0.,0.)
 \put(-5.,-40.){\epsfig{file=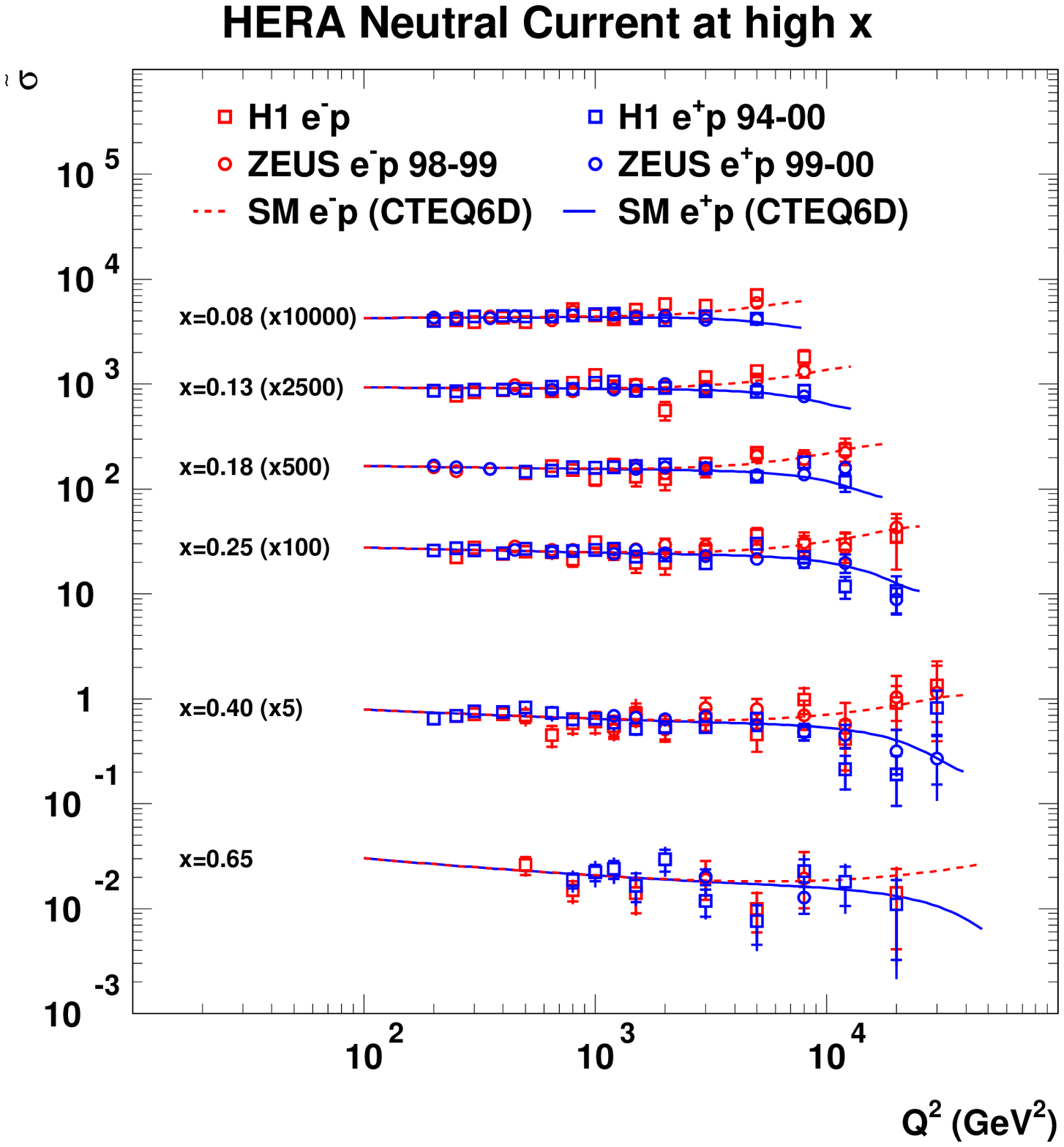,height=8.5cm}}
 \end{picture}
 \begin{picture}(170,190)(0.,0.)
 \put(30.,-40.){\epsfig{file=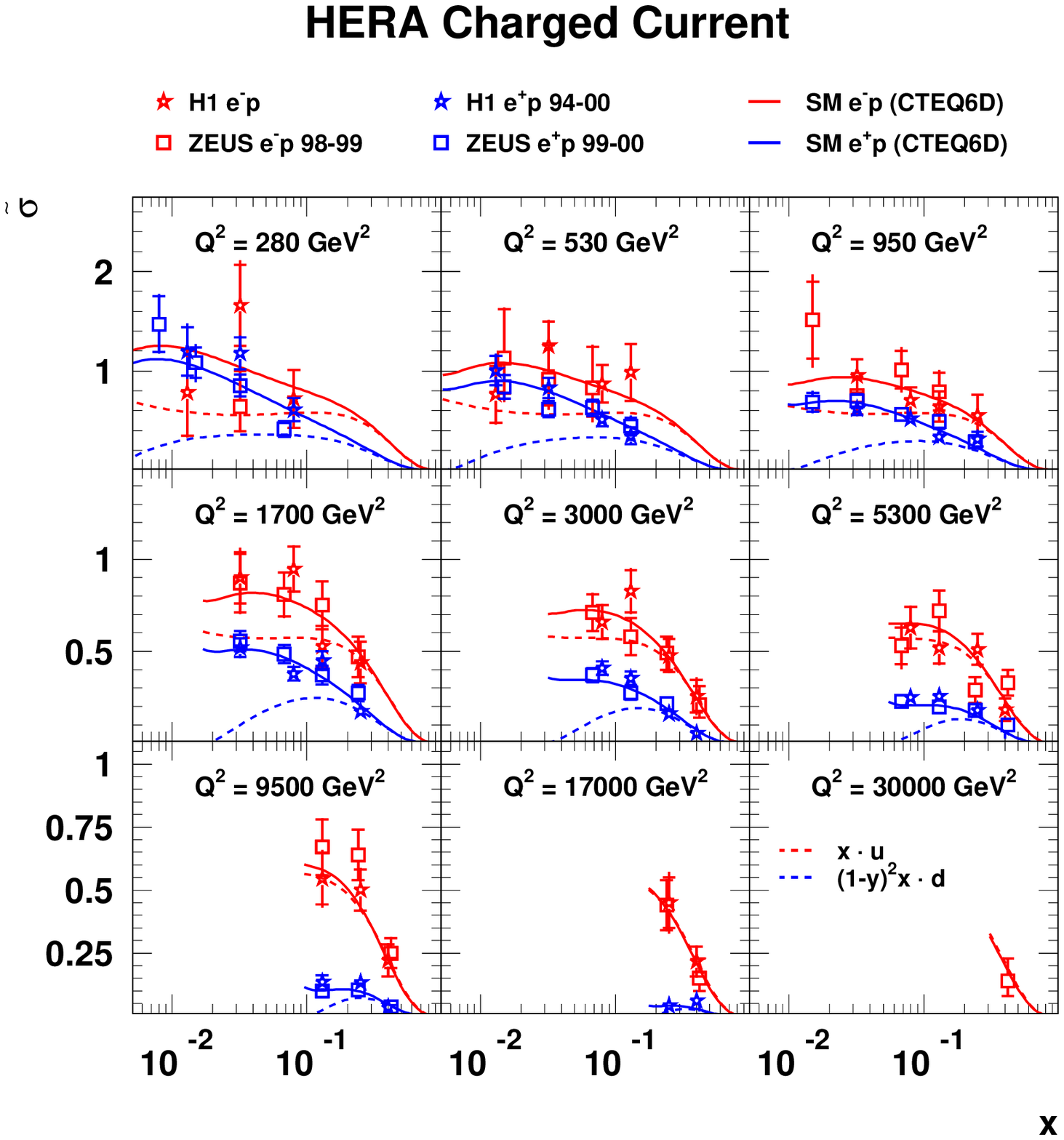,height=9cm}}
 \end{picture}
\caption{
Double differential NC (left) and CC (right) $e^{\pm}p$ cross sections
at HERA compared with the Standard Model predictions based on 
the CTEQ parton distributions~\cite{cteq6}. 
For CC, the contributions from u and d quarks are also shown.} 
\label{nccc}
\end{figure}

Double differential NC $e^{\pm}p$ cross section measurements at high $x$
are shown in Figure~\ref{nccc} (left).
At low $Q^2$ the cross sections of the $e^+p$ and $e^-p$ interactions 
are essentially indistinguishable.
At high $Q^2$ they depart from each other
due to the different sign of the $xF_3$ contribution to the cross section
(see eq.~\ref{dsignc}). 
The HERA experiments determined 
the structure function $xF_3$~\cite{h1highq2,zeuselec} 
by the difference of $e^+p$ and $e^-p$ cross sections.
It is dominated by the $\gamma-Z$ interference term
and depends on the valence quark density only.
Double differential CC $e^{\pm}p$ cross sections 
are shown in Figure~\ref{nccc} (right) as a function of $x$
for different $Q^2$.
They are sensitive to individual flavours in the proton,
the $e^+p$ data to $d$ and the $e^-p$ data to $u$ quarks.
At high $x$, the contribution from the valence quarks dominates
the cross section and allows a local extraction of $u$ and $d$
densities.

Using inclusive measurements of the NC and CC $e^{\pm}p$ cross sections,
H1 and ZEUS performed NLO QCD fits~\cite{h1highq2,zeusqcd},
which lead to a quark flavour decomposition of parton densities.
In the ZEUS fit their jet data are used as well.
The quark and gluon distributions obtained in the H1 and ZEUS fits
are shown in Figure~\ref{pdf} (left). 
The results agree within the quoted error bands.
They agree also with the parton densities
from global QCD fits~\cite{cteq6,mrst01} 
which include not only the HERA but also the fixed target
DIS data as well as data from other processes sensitive 
to parton distributions, such as Drell-Yan,
inclusive jet production and W-lepton asymmetry in $p\bar{p}$ collisions.

%%%%%%%%%%%%%%%%%%%%%%%%%%%%%%%%%%%%%%%%%%%%%%%%%%%%%%%%%%%%%%%%%%%%
%%%%%%%%%%%%%%%%%%%%%%%%%%%%%%%%%%%%%%%%%%%%%%%%%%%%%%%%%%%%%%%%%%%%
\subsection{Low-$x$ regime}

\begin{figure}[htb]
 \begin{picture}(170,180)(0.,0.)
 \put(0.,-30.){\epsfig{file=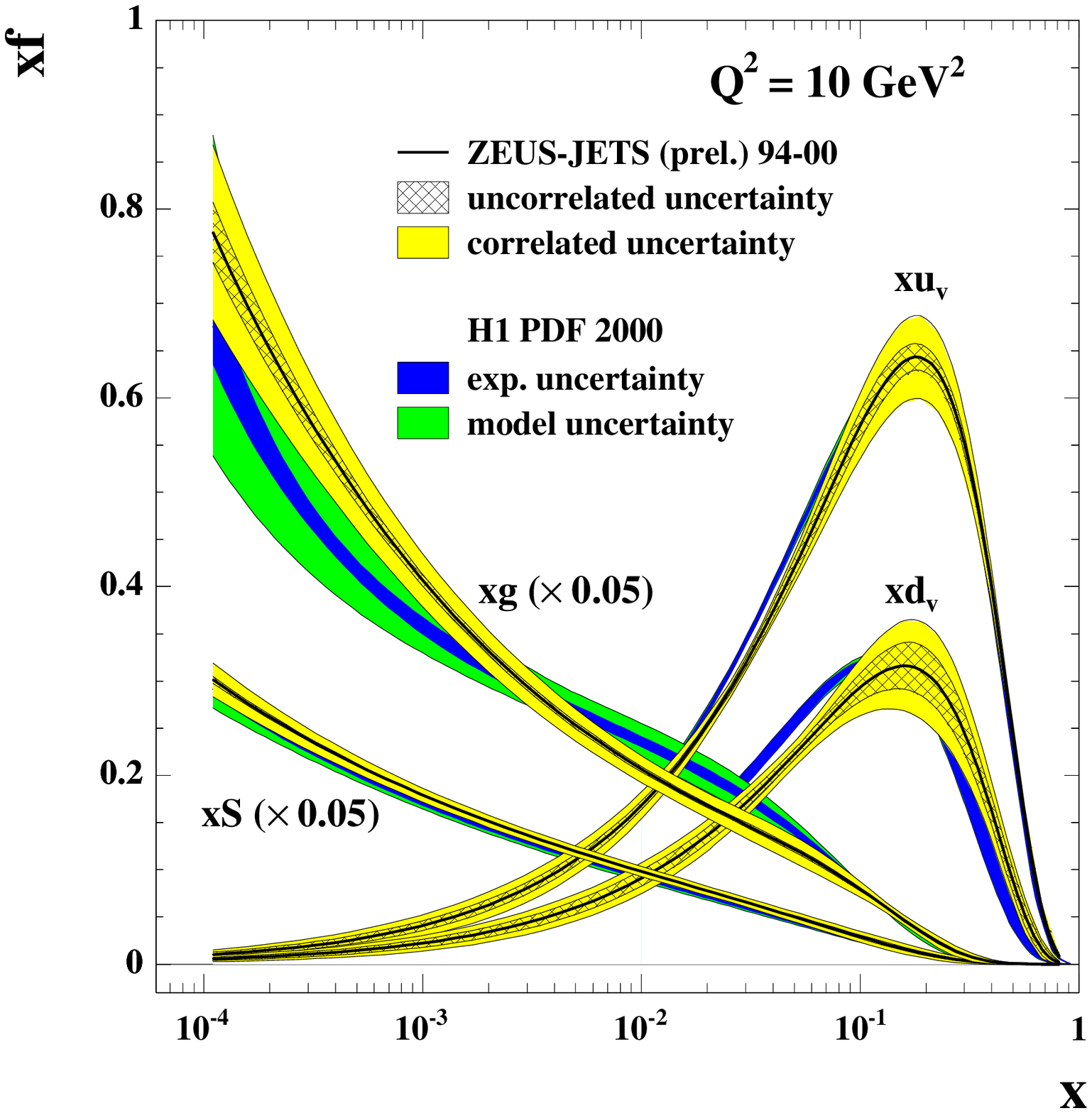,height=8cm}}
 \end{picture}
 \begin{picture}(170,180)(0.,0.)
 \put(50.,-30.){\epsfig{file=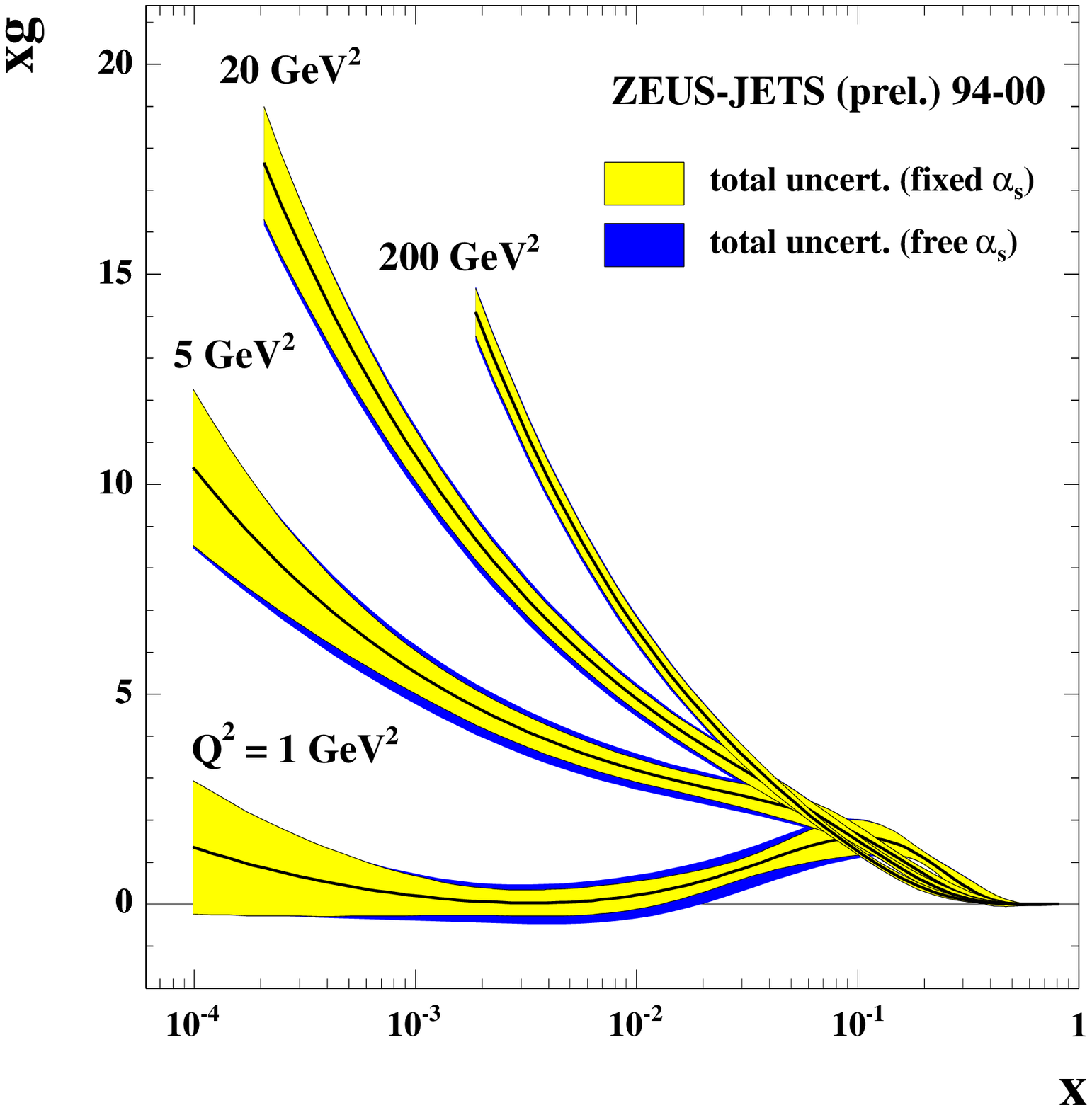,height=7.6cm}}
 \end{picture}
\caption{
The valence, sea and gluon distributions 
as obtained from the H1 and ZEUS NLO QCD fits 
to NC, CC and jet data (latter in ZEUS fit only) 
at $Q^2=10$~GeV${^2}$ as a function of $x$ (left). 
The low $x$ region is dominated by the gluon and sea quark distributions 
divided on the plot by a factor of 20. 
The gluon distribution from the ZEUS fit 
at $Q^2=$1, 5, 20 and 200~GeV$^2$ (right).
\label{pdf}}
\end{figure}

The steep rise of the proton structure function at low $x$
indicates a new partonic regime governed by a rising gluon density.
This high parton density regime might reveal novel QCD effects, 
e.g. gluon-gluon recombination.
The QCD analysis of inclusive measurements 
leads to a gluon distribution which rises 
towards small $x$ at $Q^2$ above a few GeV{$^2$}, Figure~\ref{pdf} (right).
At the lowest $Q^2$
the rise of $F_2$ persists, while
the gluon density flattens out at $Q^2\approx$~2.5~GeV{$^2$}
and even becomes dangerously close to zero at $Q^2=$~1~GeV{$^2$}.
This is not necessarily a problem in itself, since the gluon density
from scaling violations is not an observable.
A negative gluon density, however, would result 
in a distinct unphysical prediction for $F_L$.
A closer look into the low $x$ region is presented below in terms of 
derivatives of $F_2$ in $\ln x$ and $\ln Q^2$ 
and measurements of the longitudinal structure function $F_L$.

\begin{figure}[ht]
%\centerline{\epsfxsize=2.7in\epsfbox{fig7a.eps}
%\epsfxsize=2.7in\epsfbox{fig7b.eps}}
 \begin{picture}(170,160)(0.,0.)
 \put(20.,-25.){\epsfig{file=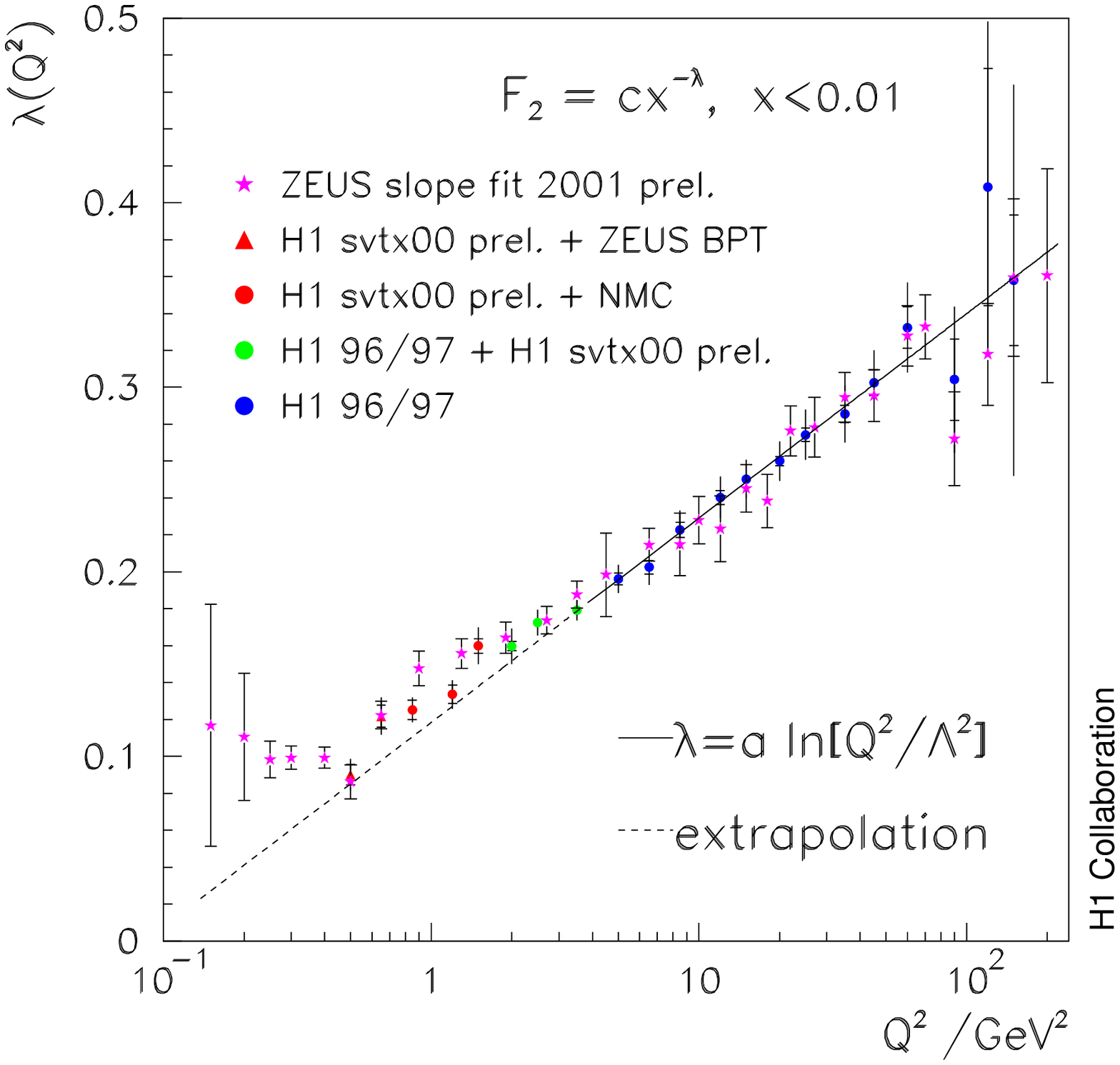,height=6.5cm}}
 \end{picture}
 \begin{picture}(170,160)(0.,0.)
 \put(60.,-23.){\epsfig{file=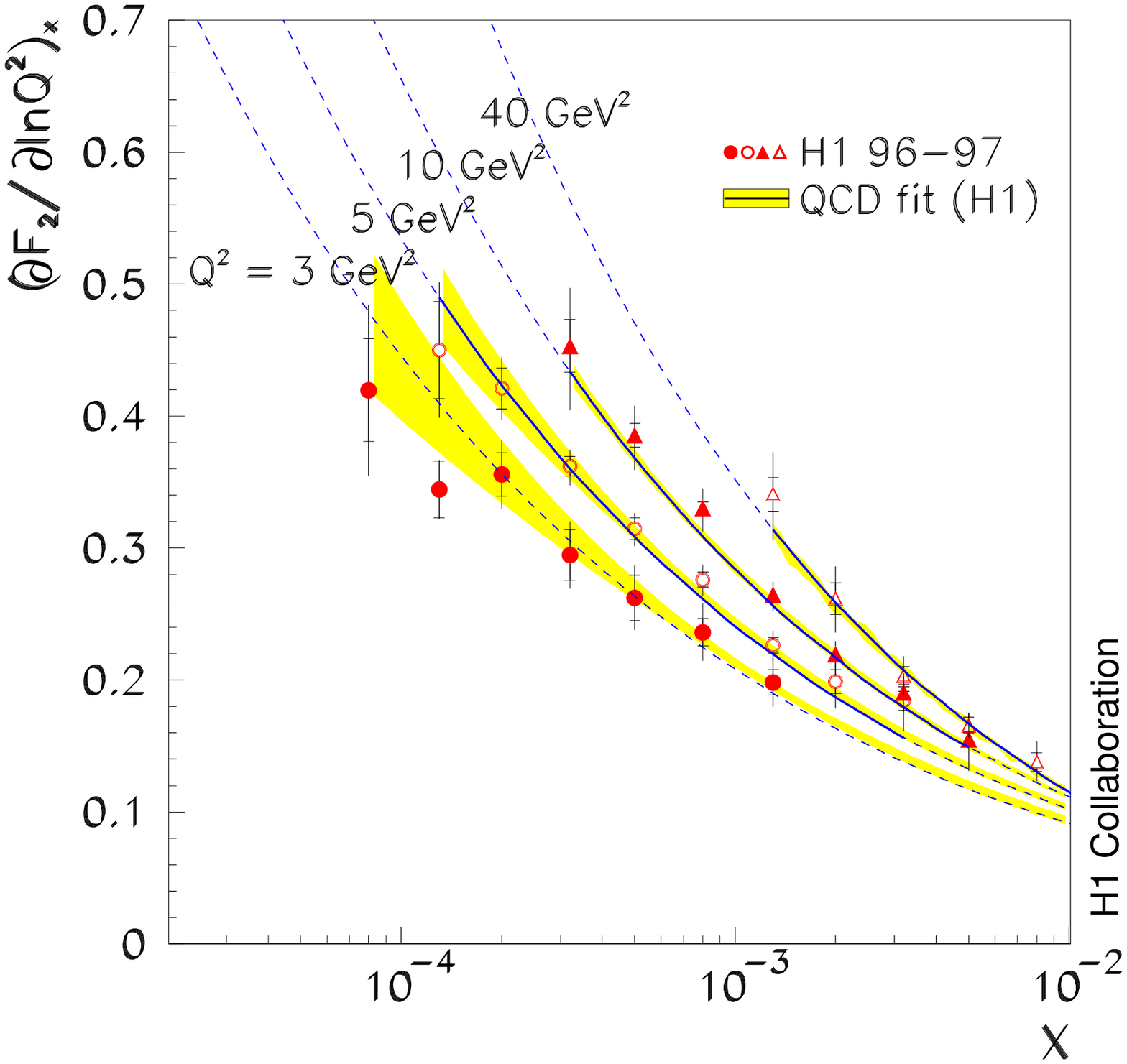,height=6.35cm}}
 \end{picture}
\caption{
Fitted values of $\lambda(Q^2)$ (right).
The derivative $(\partial F_{2} / \partial \ln  Q^{2})_x\,$
as a function of x for different $Q^2$ (right).
\label{deriv}}
\end{figure}

In the double asymptotic limit, the DGLAP evolution equation
can be solved and $F_2$ is expected to rise approximately as a power
of $x$ towards low $x$. A power behaviour is also predicted
in BFKL theory~\cite{bfkl}. A damping of this rise
would indicate the presence of novel QCD effects. 
A relevant observable
for the investigation of the dynamics of this growth is
the partial derivative of $F_2$ w.r.t. $\ln x$ at fixed $Q^2$,
$ \lambda = -(\partial \ln F_2(x,Q^2) / \partial \ln x  )_{Q^2}~$.
The high precision of the present $F_2$ data allowed H1 to measure
this observable locally~\cite{h1rise}.
The measurements are consistent with no dependence
of $\lambda$ on $x$ for $x<0.01$. 
Thus, the monotonic rise of  $F_2$ persists
down to the lowest $x$ measured at HERA, and 
no evidence for a change of this behaviour such as 
a damping of the growth is found.
The observed independence of the local derivatives in $\ln x$
at fixed $Q^2$ suggests that
$F_2$ can be parameterised in a very simple form
$ F_2 = c(Q^2) x^{-\lambda(Q^2)}~$.
The results for $\lambda(Q^2)$ obtained by H1 and ZEUS 
are shown in Figure \ref{deriv} (left). 
The coefficient $c(Q^2) \approx 0.18$
and  the parameterisation $\lambda(Q^2) = a \cdot $ln($Q^2/\Lambda^2$)
for $Q^2 \geq 2$~GeV$^2$
are consistent with pQCD analyses. 
At $Q^2 \leq 1$~GeV$^2$ the behaviour is changing, and, 
in the photoproduction limit ($Q^2\approx 0$),
$\lambda$ is approaching 0.08, which is expected from
the energy dependence of soft hadronic interactions
$\sigma_{tot} \sim s^{\alpha_P(0)-1} \approx s^{0.08}$.

Another important quantity in view of
possible non-linear gluon interaction effects is
the derivative 
$(\partial F_{2} / \partial \ln  Q^{2})_x\,$
which is a direct measure of scaling violations.
Its behaviour in $x$ 
is a reflection of the gluon density dynamics
in the associated kinematic range.
The derivative measurements are shown in Figure~\ref{deriv} (right)
as a function of $x$ for different $Q^2$.
They show a continuous growth towards low $x$
%down to lowest $Q^2 = 0.3$ GeV$^2$ (not shown), 
without an indication of a change in the dynamics.
The derivatives are well described by the pQCD calculations 
for $Q^2 \geq 3$~GeV$^2$.

\begin{figure}[htb]
 \begin{picture}(170,190)(0.,0.)
 \put(3.,0.){\epsfig{file=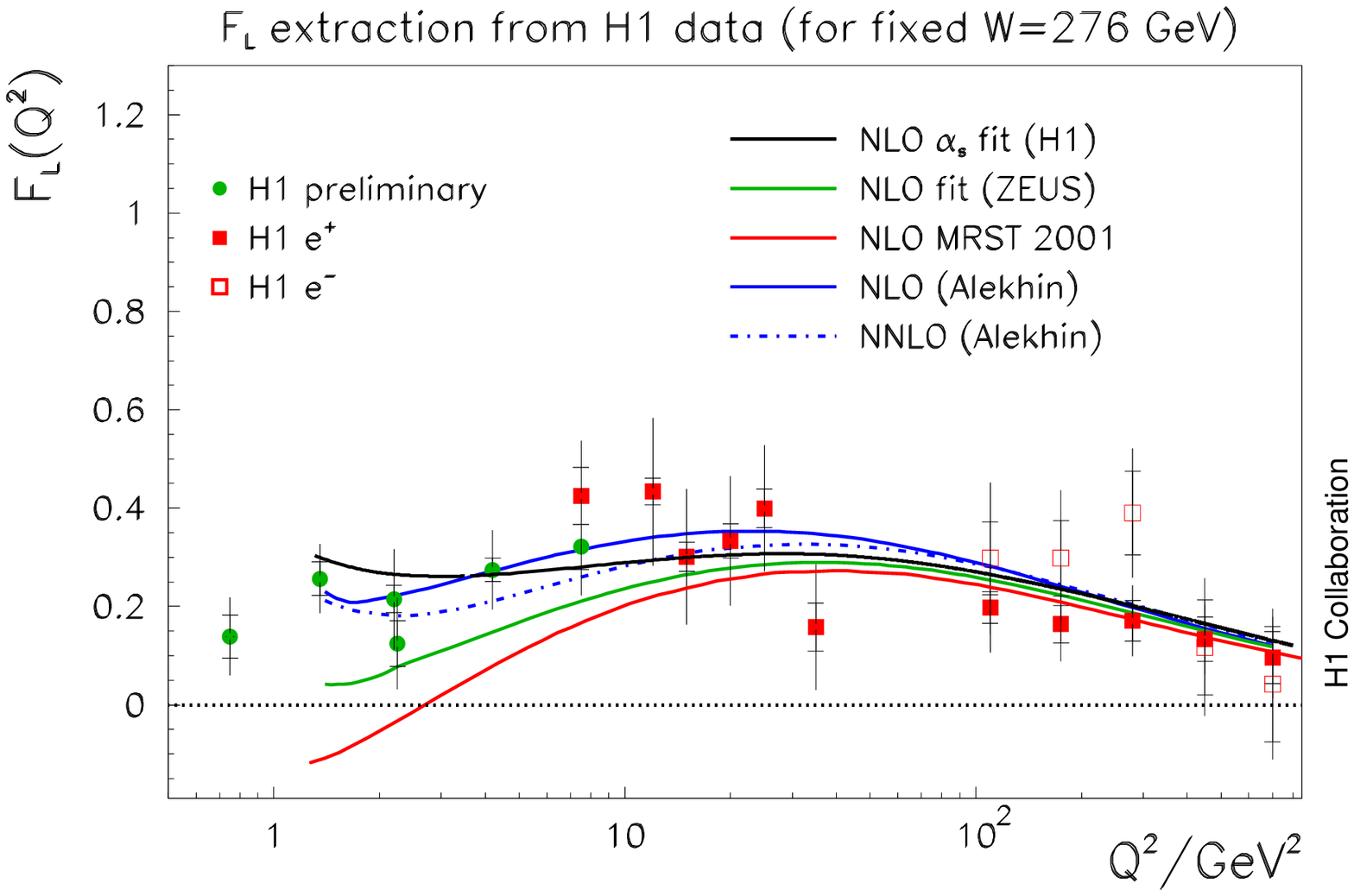,height=5.5cm}}
 \end{picture}
 \begin{picture}(170,190)(0.,0.)
 \put(30.,-50.){\epsfig{file=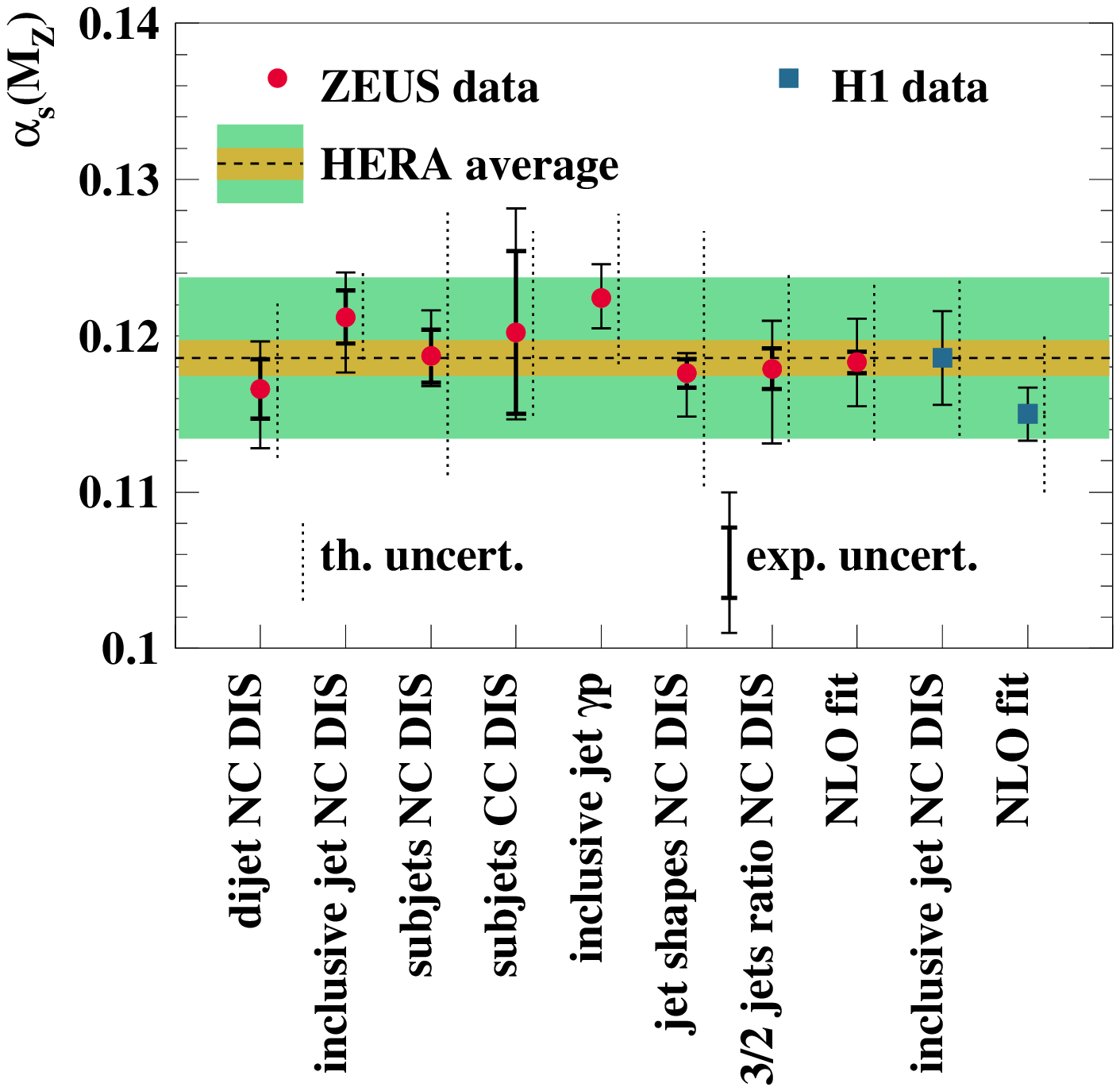,height=11cm}}
 \end{picture}
\caption{
Summary of the $F_L$ measurements by H1 
at fixed photon-proton center of mass energy 
$W=276$~GeV, $W \approx \sqrt{sy}$  (left).
Measurements of the strong coupling constant $\alpha_s(M_Z^2)\,$
in inclusive DIS and jet production at HERA (right).}
\label{flalphas}
\end{figure}

Non-zero values of the structure function $F_L$ 
appear in pQCD due to gluon radiation. 
Therefore, $F_L$ is a most appropriate quantity to test QCD to NLO 
and especially to examine pathological effects 
related to a possibly negative gluon distribution. 
According to eq.~\ref{dsignc}, the $F_L$ contribution to
the inclusive cross section is significant only at high $y$. 
The conventional way to measure $F_L$ is to explore 
the $y$  dependence of the cross section at given
$x$ and  $Q^2$ by changing the center of mass energy of the
interaction. Such measurements are not yet performed at HERA.
The H1 collaboration nevertheless could determine  $F_L$ from measurements
at high $y$, i.e. small scattered electron energies 
down to 3 GeV. Various methods are used which attribute
the observed decrease of the cross section at high $y$
to $F_L$ according to eq.~\ref{dsignc}.
A summary of the $F_L$ measurements by H1~\cite{h1fl}
is shown in Figure~\ref{flalphas} (left).
The results are significantly above zero everywhere,
including the lowest $Q^2$.
They are compared with pQCD calculations
and different phenomenological models 
showing that already at the present level of precision
the measurements can discriminate between different predictions.
Direct measurements of $F_L$ at HERA
can be performed only by reducing the beam
energy and employing the highest $y$ domain.

%%%%%%%%%%%%%%%%%%%%%%%%%%%%%%%%%%%%%%%%%%%%%%%%%%%%%%%%%%%%%%%%%%%%
\subsection{Strong coupling $\alpha_s$}

Further insight into the structure of the proton is obtained 
in semi-inclusive processes such as jet or heavy flavour production
which are calculable in pQCD.
The jet data have been included into 
the QCD analysis by ZEUS~\cite{zeusqcd}.
They help to reduce the uncertainty of the gluon distribution at medium 
$x$ around 0.1. 
In this fit both, the gluon density shown in Figure~\ref{pdf} (right)
and the fundamental coupling constant $\alpha_s(M_Z^2)\,$
were determined.  
The H1 collaboration has also performed 
a QCD fit~\cite{h1lowq2} devoted especially to the determination
of the gluon distribution and $\alpha_s$.
A novel flavour decomposition of $F_2$ used in this fit
allows to reduce the number of
parton distributions to just two combinations 
(apart from the gluon density): 
``valence-like'' and ``sea-like''.
The number of data sets can thus be reduced to a minimum,
the H1 cross section data covering low\,$x$
and the BCDMS $\mu p$ data covering the region of large $x$.
Without HERA data the gluon distribution from the fit
to the BCDMS data becomes unrealistically flat at low $x$
and drives $\alpha_s(M_Z^2)\,$ to about 0.110. The HERA data
pin down the gluon density, and, as a result, $\alpha_s(M_Z^2)\,$ moves
to 0.115 if both data sets are used.

The most precise HERA measurements 
of the strong coupling constant $\alpha_s(M_Z^2)\,$
in inclusive DIS are shown in Figure~\ref{flalphas} (right)
together with various $\alpha_s(M_Z^2)\,$ determinations
from jet production at HERA~\cite{jets}. The average $\alpha_s(M_Z^2)\,$
value in NLO from H1 and ZEUS is
\begin{equation}
  \alpha_s(M_Z^2)\, = 0.1186 \pm 0.0011(exp) \pm 0.005(theory). 
\label{alphas}
\end{equation}
The experimental error is small, about 1\%.
A limitation of the precision of $\alpha_s(M_Z^2)\,$ arises from
the associated theoretical error calculated 
using the $ad~hoc$ convention of
varying the renormalisation and factorisation scales by a factor of 2.
In forthcoming exact NNLO
analyses the scale dependence will be considerably reduced.
With theoretical and experimental progress in DIS,
it is expected to pin down the least well-measured 
fundamental constant $\alpha_s(M_Z^2)\,$ to an accuracy of better than 0.001.

%%%%%%%%%%%%%%%%%%%%%%%%%%%%%%%%%%%%%%%%%%%%%%%%%%%%%%%%%%%%%%%%%%%%
\subsection{Heavy flavour production}

The dominant heavy flavour production process at low $x$ at HERA 
is photon-gluon fusion, in which a pair of charm or beauty 
quarks and antiquarks is produced. 
The mass of the heavy quark
provides a hard scale and makes it possible to apply pQCD techniques 
for cross section calculations even in photoproduction ($Q^2 \approx 0$).
The interplay between different hard scales, mass and
$p_t$ of the heavy quark and $Q^2$, is one of the questions
which heavy flavour studies need to resolve. 

\begin{figure}[htb]
 \begin{picture}(170,200)(0.,0.)
 \put(5.,-25.){\epsfig{file=fig9a.epsi,height=8.5cm}}
 \end{picture}
 \begin{picture}(170,200)(0.,0.)
 \put(8.,25.){\epsfig{file=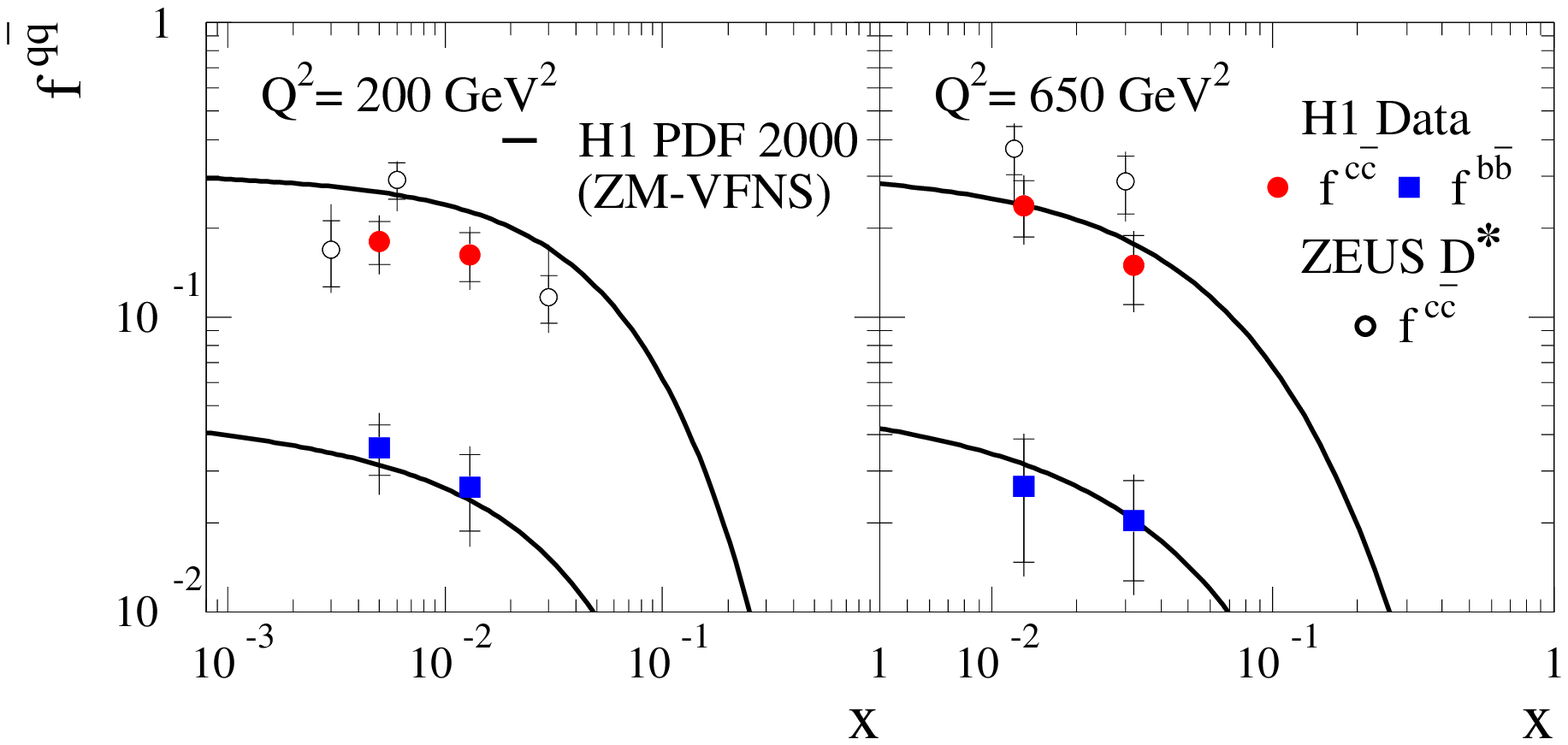,height=5.5cm}}
 \end{picture}
\caption{
Fractional contributions of charm and beauty quarks to 
the proton structure function $F_2$. The curves correspond to
predictions within the H1 and ZEUS QCD analyses.}
\label{f2ccbb}
\end{figure}

Open charm production at HERA has mainly been studied 
by reconstructing $D^*$~mesons which originate
from the hadronisation of charm quarks and decay
into $D^0\pi\rightarrow K\pi\pi$.    
%into $D^0\pi_{slow}\rightarrow K\pi\pi_{slow}$.    
The measured contribution of charm  to the proton structure 
function~\cite{h1f2cc,zeusf2cc} 
is shown in Figure~\ref{f2ccbb} (left). 
The charm contribution is large and increases from $\approx 10$\%
at $Q^2=2$~GeV${^2}$ to $\approx 30$\% at $Q^2=500$~GeV${^2}$. 
Silicon detectors surrounding the interaction region
provide lifetime tagging for heavy flavour physics at HERA.  
Using this powerful tool the H1 collaboration determined the cross sections 
for charm and beauty production~\cite{h1f2bb}
shown in Figure~\ref{f2ccbb} (right).
The vertex tagged H1 charm data are consistent with the ZEUS $D^*$ data
and complement the data obtained at lower $Q^2$.
The beauty contribution is about 1/10 of the
charm contribution and amounts to about 2\% of $F_2$.
This is the first measurement of the beauty
structure function $F_2^{b\bar{b}}$. 
The data are well described by pQCD calculations 
both for $F_2^{c\bar{c}}$ and $F_2^{b\bar{b}}$.
The accurate measurement of these structure functions
is important for the forthcoming LHC data, because their contribution
is expected to be much increased for the scale relevant for the LHC. 

%%%%%%%%%%%%%%%%%%%%%%%%%%%%%%%%%%%%%%%%%%%%%%%%%%%%%%%%%%%%%%%%%%%%
\subsection{First results from HERA II}

\begin{figure}[htb]
 \begin{picture}(170,180)(0.,0.)
 \put(110.,-40.){\epsfig{file=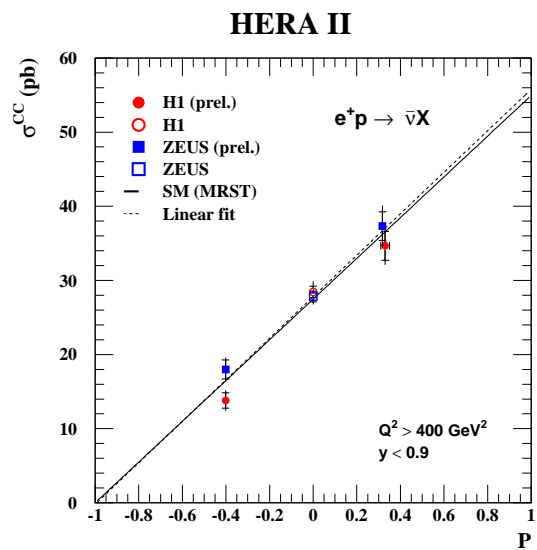,height=8cm}}
 \end{picture}
\caption{
Measurements of the total CC $e^+p$ cross section as a function of the
positron beam polarisation $P$. The published HERA I results correspond to
$P=0$. The first preliminary HERA II measurements are made
at $P=-0.40$ and $P=0.33$. The lines indicate the Standard Model expectation 
and a linear fit to the data. }
\label{hera2}
\end{figure}

Data taking of the second, high luminosity phase 
of the HERA program, HERA~II, started in October 2003. 
After the upgrade of the collider the specific luminosity 
is increased by about a factor of 3.
This was achieved by placing strong super-conducting focusing magnets inside
the H1 and ZEUS detectors, close to the interaction point.
In 2003-2004 HERA has delivered about 100 pb$^{-1}$. 
A major success at HERA II is the longitudinal polarisation
of the positron beam ($P$), typically about 40\%.
This new feature was exploited by H1 and ZEUS to measure
polarised NC and CC $e^+p$ cross sections~\cite{h1hera2,zhera2}.
The measured total CC cross sections are shown in Figure~\ref{hera2} 
together with unpolarised HERA I data ($P=0$).
The measurements confirm the predicted
linear dependence of the CC cross section on the
longitudinal polarisation of the positron beam.
The $e^+p$ cross section is expected to vanish 
at $P=-1$ unless right handed weak currents exist.
The result of the combined H1 and ZEUS cross section measurements,
linearly extrapolated to $P=-1$,
is $0.2 \pm 1.8(stat) \pm 1.6 (syst)$~pb ($\chi^2/dof=5.4/4$), 
which is consistent with absence of right handed weak currents.

%%%%%%%%%%%%%%%%%%%%%%%%%%%%%%%%%%%%%%%%%%%%%%%%%%%%%%%%%%%%%%%%%%%%
%%%%%%%%%%%%%%%%%%%%%%%%%%%%%%%%%%%%%%%%%%%%%%%%%%%%%%%%%%%%%%%%%%%%
\section{Outlook}

For more than three decades unpolarised structure function
measurements in DIS are providing crucial experimental input
to establish the quark parton model, to develop QCD
and to determine universal parton distribution
functions. The area is still very active and provides many results
from fixed target experiments and, in the past decade,
from the unique $ep$ collider HERA. The second phase of 
the HERA program with increased luminosity 
and upgraded detectors just started.
The rich physics program of HERA~II has 
the goal to collect about 1 fb$^{-1}$ of data, including
a special run with lower proton beam energies 
for direct measurements of $F_L$ at low $x$.
The end of HERA, foreseen for mid 2007,
will definitely not allow to complete the program
and will enforce compromises on the physics outcome from HERA.
The field has a large potential in a long term future.
Extensions of the present HERA program 
to study neutron structure in $ed$ collisions
and for measurements at low $x$ and $Q^2$ with improved precision
have been proposed recently~\cite{hera3}.
The HERA (or Fermilab) proton ring could serve for ep collisions 
with the future International Linear Collider at the new 
energy frontier~\cite{thera}. 
Physics opportunities and the accelerator and detector options
for a future Electron Ion Collider, EIC/eRHIC at BNL~\cite{erhic},
are under discussion.

%%%%%%%%%%%%%%%%%%%%%%%%%%%%%%%%%%%%%%%%%%%%%%%%%%%%%%%%%%%%%%%%%%%%
%%%%%%%%%%%%%%%%%%%%%%%%%%%%%%%%%%%%%%%%%%%%%%%%%%%%%%%%%%%%%%%%%%%%
%\section*{Acknowledgments}

%%%%%%%%%%%%%%%%%%%%%%%%%%%%%%%%%%%%%%%%%%%%%%%%%%%%%%%%%%%%%%%%%%%%
%%%%%%%%%%%%%%%%%%%%%%%%%%%%%%%%%%%%%%%%%%%%%%%%%%%%%%%%%%%%%%%%%%%%

\end{document}